\documentclass[12pt]{article}
\usepackage{amssymb}
\usepackage{amsfonts}
\usepackage{epsfig}
\usepackage{graphicx}

\begin{document}

\title{Reconstructing cryptocurrency processes via Markov chains}
\author{Tanya Ara\'{u}jo\thanks{ORCID 0000-0001-6993-7043} $^{a,b}$ and  Paulo Barbosa $^{a}$ \and  $^{a}$ ISEG
(Lisbon School of Economics \& Management) \and Universidade de Lisboa
\and  $^{b}$Research Unit on Complexity and Economics (UECE)}

\date{}

\maketitle

\abstract{The growing attention on cryptocurrencies has led to increasing research on digital stock markets. Approaches and tools usually applied to characterize standard stocks have been applied to the digital ones. Among these tools is the identification of processes of market fluctuations. Being interesting stochastic processes, the usual statistical methods are appropriate tools for their reconstruction. There, besides chance, the description of a behavioural component shall be present whenever a deterministic pattern is ever found. Markov approaches are at the leading edge of this endeavour. In this paper, Markov chains of orders one to eight are considered as a way to forecast the dynamics of three major cryptocurrencies.
It is accomplished using an empirical basis of intra-day returns. Besides forecasting, we investigate the existence of eventual long-memory components in each of those stochastic processes.
Results show that predictions obtained from using the empirical probabilities are better than random choices.}

\vspace{0.5cm}


\vspace{0.25cm}

JEL Classification {D8, H51}

\maketitle

\section{Introduction}

Since the introduction of Bitcoin by Nakamoto (2008), cryptocurrencies have received considerable attention from monetary authorities, firms and investors. The reasons for all this attention are the possibility of reducing risk management, improving portfolios and analyzing consumer sentiment (Dyhrberg, 2016). Indeed, few financial innovations have drawn similar attention from regulators, investors and stakeholders.

Attention naturally led to characterizing some stylized facts of the digital market. References Urquhart (2017) and Cunha, C.R. \& Silva, R. (2020) show that stylized facts of Bitcoin (\texttt{BTC}) data are similar to those of traditional financial assets. Namely, distributions of \texttt{BTC} one-day returns display fat tails, exhibit volatility clustering and the correlation between its volume and volatility is always positive. Reference Cheah et al.(2018) uses daily data to model the volatility in the cryptocurrency market. Their findings indicate that the cryptocurrencies studied (Bitcoin, Ethereum and Ripple) have long memory. Likewise, Kaya et al. (2020)  show that those cryptocurrencies have fat-tail distributions suggesting that their returns approach three standard deviations.

Simultaneously, Bariviera (2017) focus on studying \texttt{BTC} long-range memory of daily and intra-day prices. The authors show that \texttt{BTC} data display high volatility, long-range memory unrelated to market liquidity, and intra-day prices similar across different time scales (5 to 12 hours). Moreover, the persistent behavior of daily prices from 2011 to 2014 was captured by the calculation of the Hurst exponent, showing that, after 2014, it decreased to nearly $0.5$, behaving like a random process. Malladi, R.K. \& Prakash L.D. (2021), analyze the returns and volatility of \texttt{BTC} and Ripple (\texttt{XRP}). They found that returns of global stock markets and gold do not have a causal effect on \texttt{BTC} returns. However, the returns of cryptocurrencies with less market capitalization, like \texttt{XRP}, are more affected by gold prices and stock market volatility. Since the prices of \texttt{XRP} are more volatile and more affected by market news, the authors use its prices as a proxy for investor fear and show that its volatility can also drive \texttt{BTC} prices.

Likewise, references Dyhrberg (2016) and Baur (2010) show that \texttt{BTC} has similar hedging capabilities to gold against the US dollar and the Stock Exchange Index. Their arguments rely on hedging capabilities, concluding that investors can use this virtual currency alongside gold to eliminate or minimize specific market risks. In addition, as \texttt{BTC} can be traded continuously, 24 hours per day and seven days per week, Dyhrberg (2016) argues that this virtual currency has specific speed advantages and can be added to the list of hedging tools. Similarly, Deniz, A. \& Teker, D. (2020), show that \texttt{BTC} and Gold price series are cointegrated, implying that both series follow similar paths in the long term. They also show, through the Granger test, that gold prices affect \texttt{BTC} prices in a short time. However, these results are different for Ethereum (\texttt{ETH}) and \texttt{XRP} prices, where there is no direct long or/and short-term relation between these price series and gold.

Cheah et al.(2018) show that \texttt{BTC} prices behave like a traditional asset, being prices dominated by highly speculative periods. Therefore, they conclude that the cryptocurrency market shows substantial similarity in stylized empirical facts when compared to traditional markets and, more precisely, in what concerns vulnerability to speculative bubbles. In the same way, Cobert et al. (2018) also study the stochastic properties of the leading cryptocurrencies and their
linkages to standard stock market indices. The main findings show that the behavior of cryptocurrency markets is highly connected to each other but decoupled from the primary traditional stock indexes. Therefore, digital stocks are seen as an essential contribution since they offer diversification benefits to investors.

In the present paper, in order to estimate the behavior of some major cryptocurrencies, the processes of one-hour returns of \texttt{BTC}, \texttt{ETH} and \texttt{XRP} are reconstructed\footnote{The Ethereum platform was launched in 2015, being best known by its cryptocurrency and token ETH. In the short term, ETH and BTC have many similarities but different long-term constraints. Similarly, the cryptocurrency XRP is owned by Ripple, which was first released in 2012, being a blockchain-based digital payment protocol. The main goal of XRP is to serve as an intermediate device of exchange between two currencies, acting as a sort of temporary liquidation channel.}.
It is done by using Markov chains of orders one to eight.
The reconstruction of those cryptocurrencies processes starts with the identification of: \textit{(i)} the allowed (markovian) transitions in the state space that corresponds to current orbits of the system, and \textit{(ii)} the occurrence frequency of each orbit in typical samples. It is accomplished by taking the first half ( the past) of each sample to the computation of the conditional probabilities of the allowed transitions. From this empirical base, each second half (the future) is estimated and compared with a random choice.  Results show that the predictions obtained from the empirical transitions probabilities outperform random forecasts.

A short overview of the subsequent sections of this paper consists in the following: the next section describes the data used in our empirical approach. Such a description is followed by the presentation of the methodology: a Markov chain model that underlies the reconstruction of three cryptocurrency processes. Section 3 presents and discuss the results obtained from such a reconstruction and compare them with random forecasts. The last section presents and concluding remarks and outlines future work.

\section{Data and Methods}\label{sec11}

\begin{figure}[h!]
  \centering
  \includegraphics[width=.8\textwidth]{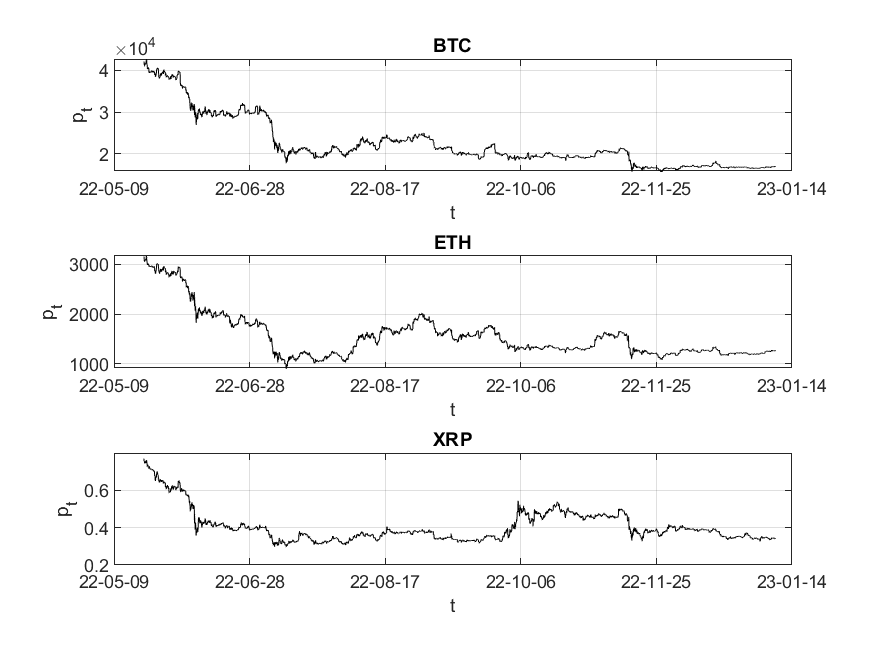}
  \caption{Hourly-prices $p_t$ of the three cryptocurrencies, from May-20-2022 to Jan-08-2023.}
   \label{fig1}
\end{figure}

The data used in the present study is sourced from \texttt{coinmarket.com}, where historical information about over one thousand cryptocurrencies is available.
Three major cryptocurrencies according to their market capitalization: \texttt{BTC}, \texttt{ETH} and \texttt{XRP} were considered. Fig. \ref{fig1} shows the behavior of hourly-price data for those three digital stocks.

The three plots in Fig.\ref{fig1} present the series of hourly-prices of \texttt{BTC}, \texttt{ETH} and \texttt {XRP}, from May-20-2022 to Jan-08-2023, comprising each of them 6306 observations.

From the plots in Fig.\ref{fig1}, it is quite apparent that the data are very far from stationary, but a different situation comes out when, instead of  prices, we considered the series of one-hour returns.

\begin{figure}[h!]
  \centering
  \includegraphics[width=.8\textwidth]{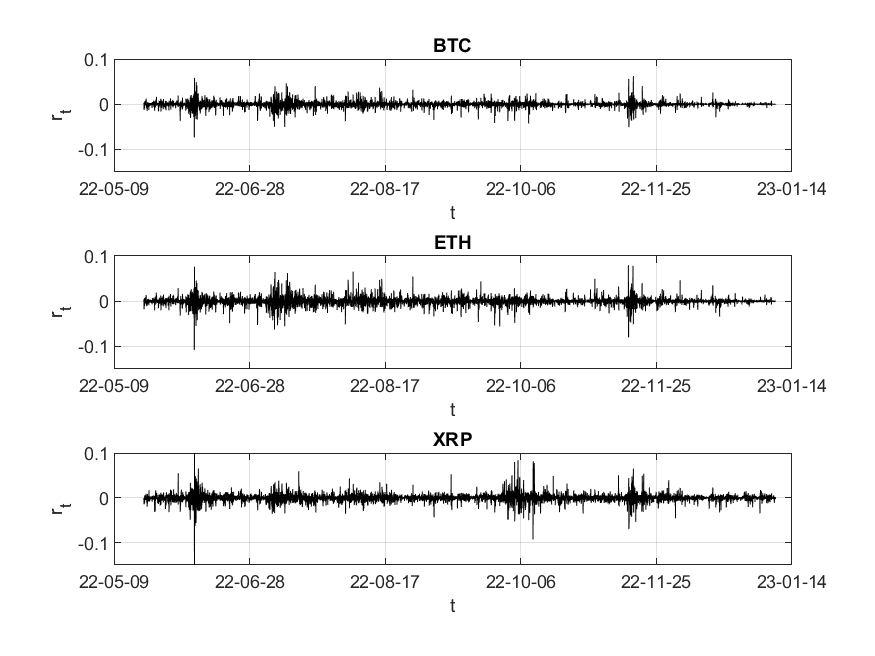}
  \caption{Hourly-returns of  the cryptocurrencies, from 2022-05-20 to 2023-01-08}
  \label{fig2}
\end{figure}

\begin{equation}
\label{eq1}
r_{t} = log(p_{t}) - log(p_{t-1}).
\end{equation}

\begin{figure}
  \centering
  \includegraphics[width=.7\textwidth]{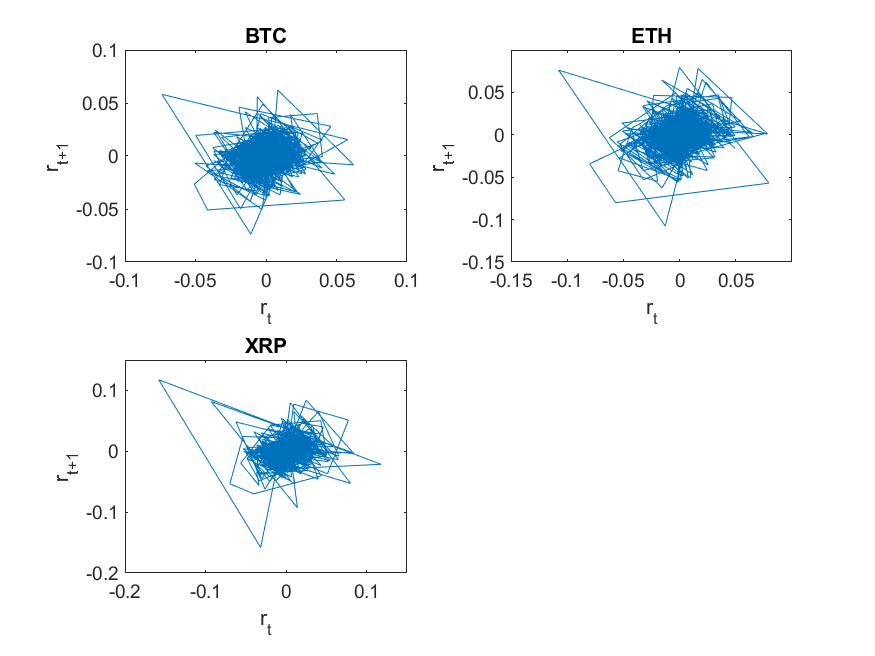}
  \caption{The dynamics of hourly-returns of the three cryptocurrencies, i.e., the phase spaces of those dynamical systems, as Eq.\ref{eq2} formally states.}
  \label{fig3}
\end{figure}

The three plots in Fig.\ref{fig2}, show the series the one-hour returns, while the plots in Fig.\ref{fig3} illustrate their dynamics, i.e., the phase spaces of those dynamical systems, as Eq.\ref{eq2} formally states.

\begin{equation}
r\left( t,1\right) \rightarrow r\left( t+1,1\right)  \label{eq2}
\end{equation}

These last three plots (Fig.\ref{fig3})  show that, in the three cases, the bulk of the data consists of a central core of small
fluctuations with a few large flights away from the core. This structure of
the data will have influence on the results obtained in the next
section.

\subsection{Markov Chains}

There is a huge literature applying Markov Chains to model the behaviour of financial time series. This approach is a fundamental tool in the study of stochastic processes. It has been used widely in many different disciplines, like weather, epidemic, land use, consumer behavior and even for the identification of writers (Khmelev, D.V., \& Tweedie, F. J. (2001)).

A sequence of random variables \(Z_1, Z_2, \ldots, Z_t, \ldots\) with Markov characteristics is known as a process with first order dependence (as described in Equation \ref{eq3}, this process has no memory). The Markov characteristics means that the distribution of the future realization of \(Z_{n+1}\) depends on its immediately previous state (\(Z_n\)) and not on further previous states (\(Z_{n-1}, Z_{n-2}, \ldots\)). Formally,

\begin{equation}
%\begin{split}
\label{eq3}
Pr(Z_{n+1} = z_{n+1} \mid Z_1 = z_1, \ldots, Z_n = z_n) \\  =
Pr(Z_{n+1} = z_{n+1} \mid Z_n = z_n)
%\end{split}
\end{equation}

In a dynamical system, shifting from state $i$ to state $j$ has transition probability $ p_{ij} $

\begin{equation}
\label{eq4}
p_{ij} = Pr(Z_{1} = s_{j} \mid Z_{0}=s_i)
\end{equation}
%.............
\vspace{0.1cm}

Fortunately, Markov chains can be approached from a higher order perspective, being Markov chains of higher orders the processes in which the next state depends on two or more preceding ones. Here, Markov chains of orders one to eight are considered as a way to predict cryptocurrency hourly-returns and to investigate the existence of eventual long-memory components in that stochastic process.

\subsection{Coding}

We consider the dynamical system being coded by a finite alphabet $\Sigma $. Then, $\Omega $, the space of orbits of the system are comprised of infinite sequences $
\omega =i_{1}i_{2}\cdots i_{k}\cdots $, $i_{k} \in \Sigma $, with the
dynamical law being a shift $\sigma $ on these symbol sequences.
\begin{equation}
\sigma \omega =i_{2}\cdots i_{k}\cdots  \label{eq5}
\end{equation}
Depending on the dynamical law of the coded system, not all sequences will
be allowed. The set of allowed sequences in $\Omega $ defines the {\it %
grammar} of the shift. The set of all sequences which coincide on the first $%
n$ symbols is called a  $n-$block and is denoted $%
[i_{1}i_{2}\cdots i_{n}]$. The probability measures over the  $n-$blocks constitute part of information that may be inferred from the data, being
the main tool used to characterize the dynamical properties of the dynamical system.

%.............
\vspace{0.2cm}

To calculate the probability measures over the $n-$blocks the following computation is performed: for each series of one-hour returns ($r_t$) with mean $\hat r$ and standard deviation $s$, a five-symbols code $\Sigma$ is defined.

\begin{equation}
\Sigma =\left\{ -2,-1,0,1,2\right\}  \label{6}
\end{equation}

 Then,
\begin{equation}
\begin{array}{clc}
\left( r\left( t\right) -\overline{r\left( t\right) }\right) >s &
\Longleftrightarrow & 2 \\
s\geq \left( r\left( t\right) -\overline{r\left( t\right) }\right) >\frac{s}{%
3} & \Longleftrightarrow & 1 \\
\frac{s}{3}\geq \left( r\left( t\right) -\overline{r\left( t\right) }\right)
>-\frac{s}{3} & \Longleftrightarrow & 0 \\
-\frac{s}{3}\geq \sigma \left( r\left( t\right) -\overline{r\left( t\right) }%
\right) >-s & \Longleftrightarrow & -1 \\
-s\geq \left( r\left( t\right) -\overline{r\left( t\right) }\right) &
\Longleftrightarrow & -2
\end{array}
\label{7}
\end{equation}

This coding is used and the empirical frequencies $\widetilde{\mu }\left(
[i_{1}\cdots i_{k}]\right) $ for blocks of successively larger order $k$ are
found. Naturally, $k$ cannot be arbitrarily large because of statistics. The reliability of
results is threatened whenever $5^k$ is larger than the size $N$ of the data sample. Therefore, statistical reliability may be directly tested
by comparing the number of different occurring blocks and $5^{k}$.

\begin{figure}[h!]
     \centering
\includegraphics[width=0.8\textwidth]{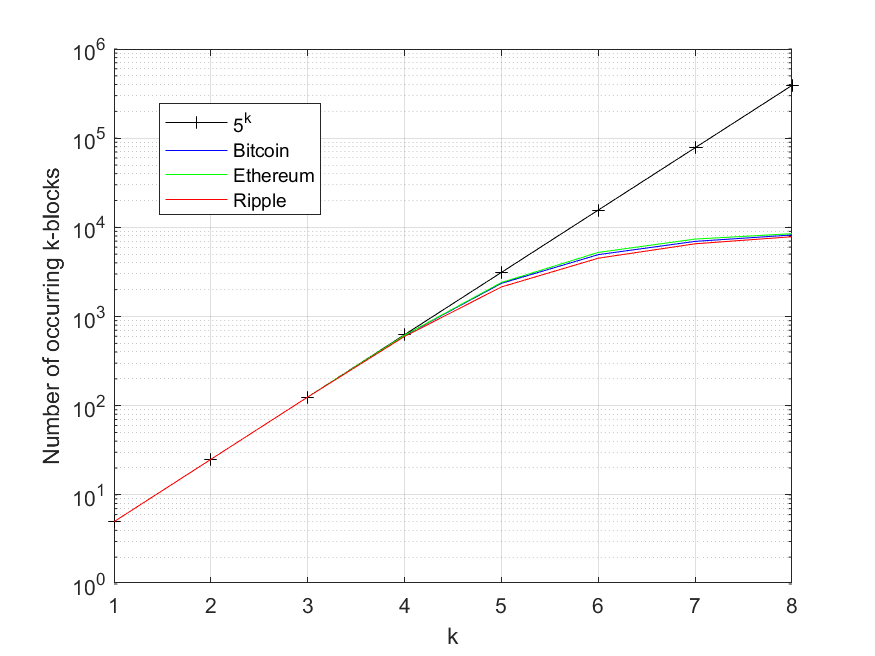}
\caption{Comparing the number of different occurring blocks of size $k$ and $5^{k}$}
  \label{fig4}
\end{figure}

Fig.\ref{fig4} shows the evolution of the number of occurring $k-$blocks and $5^{k}$, where
the number $p\left(k\right) $ of occurring blocks of size $k$ in each data sample is compared with the maximum
possible number, $5^{k}$. In all cases, after $k=4$ the comparison
 shows the lack of statistics apparent in the comparison
of $p\left( k\right) $ with $5^{k}$. These results seem to suggest that the data are described by a short range memory.

\subsection{Predicting}

Prediction starts by taking the time series of each cryptocurrency and spliting the series into two halves, the first half is then used to predict the other half.
The coding procedure presented in the last section is used and after the first half of each sample is coded into $\Sigma =\left\{ -2,-1,0,1,2\right\}$, we perform the computation of the conditional probabilities of the allowed transitions on these symbol sequences. The conditional probabilities are computed for blocks of successively larger order ($k$).

As an example, the conditional probabilities computed for blocks of size two
are defined in the following Markov transition matrix.

\[
  P =
  \left[ {\begin{array}{ccccc}
    p_{11}  & p_{12} & p_{13} & p_{14} & p_{15} \\
    p_{21} & p_{22} & p_{23} & p_{24} & p_{25} \\
    p_{31} & p_{32} & p_{33} & p_{34} & p_{35} \\
    p_{41} & p_{42} & p_{43} & p_{44} & p_{45} \\
    p_{51} & p_{52} & p_{53} & p_{54} & p_{55} \\
  \end{array} } \right]
\]

where each $p_{ij}$ indicates the probability of shifting from state $i$ to state $j$, as in Equation \ref{eq3}.

The transition matrices for blocks of orders up to eight are computed. From this empirical base, each second half is estimated and compared with a random number chosen at random.

However, when the conditional probabilities are inferred from limited experimental data an extended Markov approximation is
more convenient, we used the Less-than-$k$-Markov approach as defined by Vilela-Mendes et al. (2002).

\subsubsection{The Less-than-$k$-Markov approach}

In each simulation, with an approximation of order $k$, we look at the
current block $\left( a_{1}\cdots a_{k}\right) $ of order $k$ and use the $k-$empirical probability to infer the next state $a_{0}$. If that block is not present in the data that were used to construct the empirical probabilities, then we look at the $k-1$ sized block $a_{1}\cdots a_{k-1}$ and use the $k-1$ order empirical probabilities. If required, the process is repeated until an available empirical probability given by a $(k-i)-$order block with $2<i<8$ is found.

Such an extended Markov approach is applied to each  $k-$order block in order to estimate the successor $a_{0}$ of each block $\left(
a_{1}\cdots a_{k}\right)$. In so doing,  the successor $a_{0}$ is compared with a prediction $\widetilde{a}_{0}$
obtained by throwing a random number with the empirical probabilities
$ \widetilde{P}\left( a_{0} \mid a_{1} \cdots a_{k}\right)$.

\subsubsection*{The Past predicting the future}
Once the empirical probabilities are computed from the first half of each series ($t=1,2,...n$)
the second half is visited ($t=n+1,n+2,...2n$) in order to quantify the magnitude of the error found when using each $k-$sized block, the quantity $e_k(t)$ is computed for each sample: \texttt{BTC}, \texttt{ETH} and \texttt{XRP}.

\begin{equation}
e_k(t)= [\widetilde{a}_{0}-a_{0}(t)]
\label{eq8}
\end{equation}

As half of each series comprises $n$ observations, the averaged error for each $k-block$ is calculated

\begin{equation}
e_k =\frac{1}{n} \sum_{t=n+1}^{2n} [\widetilde{a}_0-a_0(t)]
\label{eq9}\end{equation}

The average error of the forecast of the second half of each sample is therefore computed as the distance between the observed $a_0$ and the corresponding estimated state $\widetilde{a}_{0}$.

The same procedure is performed with the successor $a_{0}$ of each $k-$order block being estimated at random ($\widetilde{r_0}$). There, the error is given by the distance

\begin{equation}
eRand_k =\frac{1}{n}  \sum_{t=n+1}^{2n} [\widetilde{r_0}-a_0(t)]
\label{eq10}\end{equation}

In the end, the errors $e_k$ and $eRand_k$ are averaged over 50 different runs.

\subsection{Method outline}

In the following, a brief and summarized description of the algorithm used in the simulations is presented. The final results contain average values over 50 runs for each cryptocurrency.

\vspace{0.2cm}

\begin{center}
\vspace{.2cm}
\textbf{Outline of the algorithm}
%\vspace{.2cm}
\begin{description}
\item Take each cryptocurrency series of hourly-prices: \texttt{BTC}, \texttt{ETH} or \texttt{XRP}
\begin{enumerate}
\item Compute hourly-returns:  $r_{t} = log(p_{t}) - log(p_{t-1})$ from the series of prices $p_t$ (Eq.1)
\item Split the series of returns into two halves of the same size $n$: $H_1$ and $H_2$
\item Code the two halves $H1$ and $H2$ into $\Sigma =\left\{ -2,-1,0,1,2\right\}$ (Eq.7)
\item Perform the steps (4.1 to 4.5) along $j=1,2,...,50$ simulations:
\begin{enumerate}
\item [] For each $k-block$ successively larger $k=2,...,8$
\begin{enumerate}
\item []4.1 Build the conditional probabilities $\widetilde{P}[( a_0[a_1\cdots a_k])$
\item []4.2 Look at the $k$ sized block $a_{1}\cdots a_{k}$ and use the $k$ order empirical probabilities to infer each next state $a_0$ with
\item []$\Diamond$  If the block $a_1\cdots a_k$ is not found, repeat using blocks of size $(k-\i)$ with $2<i<k$  until the available empirical probability is found
\item []4.3  Visit each ${a}_0(t) \in H2$, $t=n+1,n+2,...2n$
\item[]4.4 Measure the error $e_k(j)=\frac{1}{n}$$(\sum_{t=n+1}^{2n}[a_0(t)-\widetilde{a}_0 ]$) (Eq.8)
\item []4.5 Compute $eRand_k(j)=\frac{1}{n}(\sum_{t=n+1}^{2n}[a_0(t)-r_0]$ ) being $r_0$ chosen at random (Eq.9)
\end{enumerate}\end{enumerate}
\item Compute the average errors $e_k$  and $eRand_k$ over $j=1,2,...,50$ simulations (Eq.10)
\begin{enumerate}
\item []$\bullet$ $e_k =\frac{1}{50}\sum_{j=1}^{50} e_k(j)$
\item []$\bullet$ $eRand_k=\frac{1}{50}(\sum_{j=1}^{50} eRand_k(j)$)
\end{enumerate}\end{enumerate}
\end{description}
\end{center}

\section{Results and Discussion}

\begin{figure}[h!]
     \centering
\includegraphics[width=0.9\textwidth]{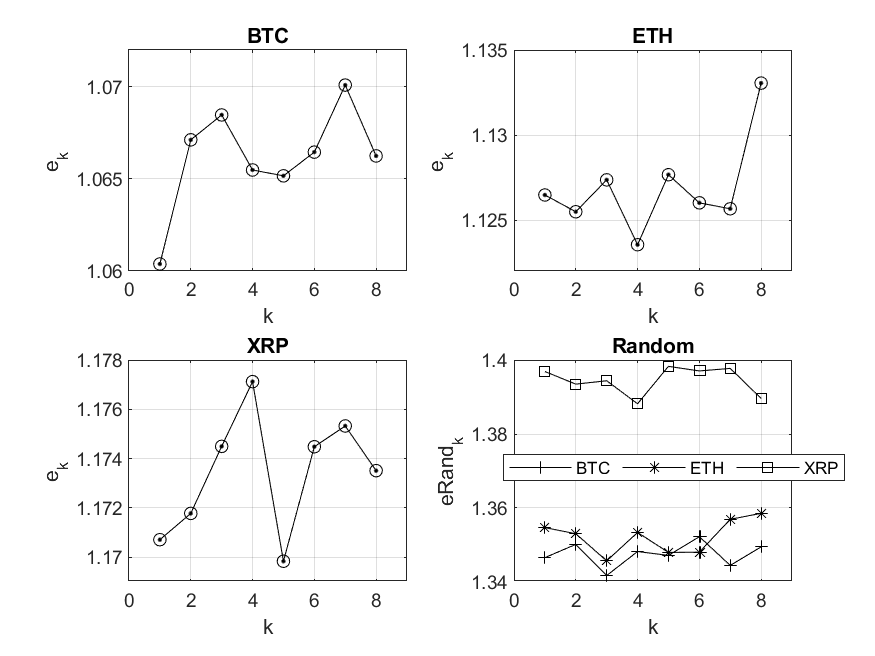}
  \caption{The first half of each sample (the past) predicting the second half (the future), using a 5-symbols alphabet.}
  \label{fig6}
\end{figure}

The first three plots in Fig.   \ref{fig6}, show the average error obtained with a 5-symbols alphabet for the three cryptocurrencies. The last plot shows the error obtained for each cryptocurrency and computed when the prediction is performed at random, i.e., from a surrogate matrix of probabilities.

These results are similar to those obtained by Vilela-Mendes et al. (2002) where daily returns of three standard stocks and the NYSE index were analysed.
Not surprisingly, in all cases, the average prediction obtained from using the empirical probabilities is better than a random choice.
However, here, \texttt{ETH} and \texttt{XRP} data show even higher improvements coming from the four and five-symbol blocks, respectively.

The stocks \texttt{BTC} and \texttt{ETH} seem to share closer similarities than any of them with  \texttt{XRP}. However, the one-symbol probabilities are similar in the three  digital stocks. This suggests that the statistical short-memory component of the market process might be similar for many different stocks, whether for the long-memory component such a similarity might be lost.

As already mentioned in the previous section, statistical limitations underlie the improvements obtained by using higher order blocks, showing much larger fluctuations.

Results previously obtained (Vilela-Mendes et al., 2002) and the need to further characterize the presence of small and large fluctuations, led to the application of the same method, with same data samples being coded by a 3-symbol alphabet. As before, $s$ is the standard deviation of the hourly-returns samples.

\begin{equation}
\Sigma =\left\{-1,0,1\right\}  \label{11}
\end{equation}

 Then,
\begin{equation}
\begin{array}{clc}
\left( r\left( t\right) -\overline{r\left( t\right) }\right) >s &
\Longleftrightarrow & 1 \\
s\geq \left( r\left( t\right) -\overline{r\left( t\right) }\right)
>s & \Longleftrightarrow & 0 \\
-s\geq \left( r\left( t\right) -\overline{r\left( t\right) }\right) &
\Longleftrightarrow & -1
\end{array}
\label{eq12}
\end{equation}

When this shorter code is adopted, the number of large events is the same as before being the statistics of small fluctuations improved.
The method is the same with the single replacement of the 5-symbol alphabet by the new one $\Sigma =\{-1,0,1\}$ with just three symbols.

\begin{figure}[h!]
     \centering
\includegraphics[width=0.9\textwidth]{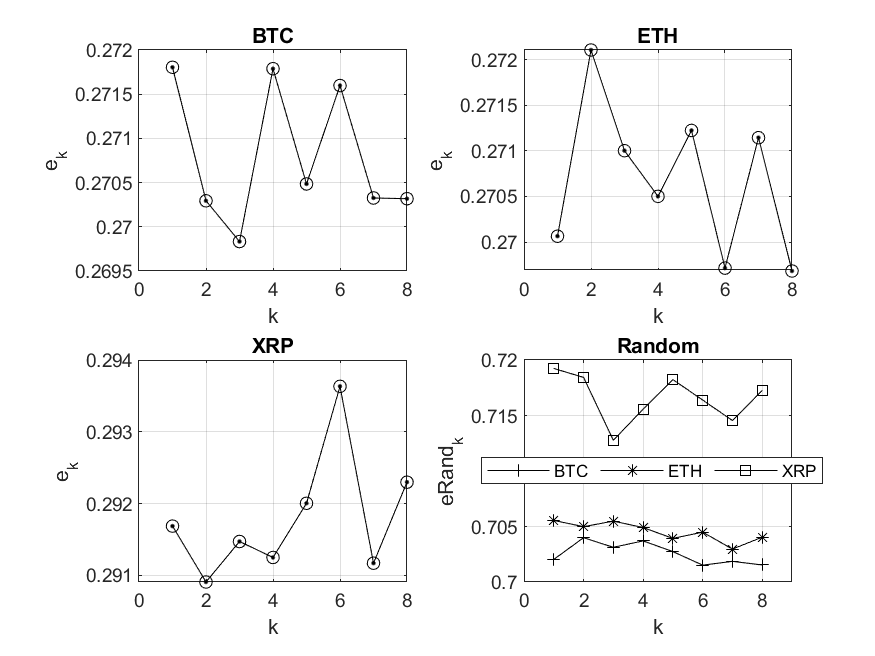}
  \caption{The first half of each sample (the past) predicting the second half (the future), using a 3-symbols alphabet.}
  \label{fig7}
\end{figure}

The first three plots in Fig.\ref{fig7} show the average error obtained with a 3-symbols alphabet for the three cryptocurrencies. The last plot shows the error obtained for each cryptocurrency and computed when the prediction is performed at random, i.e., from a surrogate matrix of probabilities.

Results show a prediction improvement extending to block sizes larger than before (with the 5-symbol alphabet). Because small fluctuation errors are decreased by better statistics, the persistence of the improvement for larger blocks seems to highlight the presence of a long-memory component.
Again, the stocks \texttt{BTC} and \texttt{ETH} seem to share closer similarities than any of them with \texttt{XRP}.

The first two plots  Fig.\ref{fig7}  show that $e_k$ computed for \texttt{BTC} and \texttt{ETH} is equally ranged in the y-axis. On the contrary,  $e_k$ computed for \texttt{XRP} displays quite different limits. A closer correlation between the first two stocks is also present in the values of $eRand_k$ as the last plots in Fig.\ref{fig6} and Fig.\ref{fig7} show.

These difference displayed by \texttt{XRP} fluctuations is certainly related to the much larger flights observed in the dynamics of \texttt{XRP} hourly returns presented in Fig.\ref{fig3}.
The improvement in the predictions obtained for small-order blocks is similar to those presented in reference Vilela-Mendes et al. (2002), where the dynamics of standard stocks was analyzed. A similar approach has
also analyzed the dependence of memory on the dynamics of processes of cryptocurrency daily-returns.
There, reference (Nascimento, 2022) - also looking at Bitcoin, Ethereum and Ripple - found the occurrence of long-range memory up to 7-order Markov chains.

\section{Conclusions}

In this paper, a Markov approach is used to model the fluctuations of processes of hourly-returns of three cryptocurrencies: Bitcoin, Ethereum and Ripple. Markov chains of orders one to eight were considered as a way to predict cryptocurrency returns and to investigate the occurrence of eventual long-memory components in those stochastic processes.

Since conditional probabilities are inferred from limited experimental data, an extended Markov approximation seems to be advantageous. Here, we used the Less-than-$k$-Markov approximation as presented in reference Vilela-Mendes et al. (2002).

The most important result is that the average predictions obtained from using the empirical probabilities outperform a random choice.

The main contributions rely on a predictive approach not yet used for series of cryptocurrencies. Moreover, using hourly data we benefit from better statistics when compared with daily ones but avoiding
the inconvenient of high-frequency data (i.e. minute observations) since it involves the interplay of many more reaction time scales and market compositions in the trading process. Therefore, the choice of series of hourly observations seems to be an appropriate way to understand the stochastic process that underlies the market mechanism.

Notice, however, that the trade-off between higher order approximations and the lack of statistics is the main
 limitation of our approach.
 Future work is planned to apply the same approach to explore the use of the empirical probabilities of one cryptocurrency to predict the behavior of the others. In so doing, we would be able to understand the strength of connectivity between digital stocks in the behaviour of the cryptocurrency market.

\vspace{0.5cm}

\textbf{{Acknowledgments}}

\vspace{0.5cm}

This article is part of the Strategic Project UIDB\/05069\/2020. The authors acknowledge financial Support from FCT –  Funda\c{c}\~{a}o para a Ci\^{e}ncia e Tecnologia (Portugal).

\section*{References}

%%%%%%%%%%%%%%
% References %
%%%%%%%%%%%%%%

\begin{itemize}
\item [] Bariviera, A.F., Basgall, M.J., Hasperue, W. \&
Naiouf, M. (2017). Some Stylized Facts of the Bitcoin Market. \emph{Physica A:
Statistical Mechanics and Its Applications} 484: 82–90.\\
https://doi.org/10.1016/j.econlet.2015.02.029.

\item [] Baur, D. \& Lucey, B.M. (2010). Is Gold a Hedge or a Safe Haven?
An Analysis of Stocks, Bonds and Gold. \emph{Financial Review} 45 (2): 217–29.\\
https://doi.org/10.1111/j.1540-6288.2010.00244.x.

\item [] Cheah, E.T., Mishra, T., Parhi, M. \& Zhang, Z. (2018), Long memory interdependency
and inefficiency in bitcoin markets, \emph{Economics Letters}, 167, 18–25.

\item [] Corbet, S., Meegan, A., Larkin, C., Lucey, B. \& Yarovaya, L. (2018), Exploring the
dynamic relationships between cryptocurrencies and other financial assets,\emph{ Economics
Letters}, 165, 28–34.\\
https://doi.org/10.1016/j.econlet.2018.01.00

\item [] Cunha, C.R. \& Silva, R. (2020). Relevant Stylized Facts about Bitcoin:
Fluctuations, First Return Probability, and Natural Phenomena. \emph{Physica
A: Statistical Mechanics and Its Applications}, 550: 124–55.
https://doi.org/10.1016/j.physa.2020.124155.

\item [] Dyhrberg, A.H. (2016). Hedging Capabilities of Bitcoin. Is It the
Virtual Gold? Finance Research Letters 16: 139–44.\\
https://doi.org/10.1016/j.frl.2015.10.025.

\item [] Kaya, P., Okur, M., Ozgur C., \& Altintig, Z. A. (2020),
2020. Long Memory in the Volatility of Selected Cryptocurrencies:
Bitcoin, Ethereum and Ripple. \emph{Journal of Risk and Financial Management}
13 (6): 107.

\item [] Khmelev, D.V. \& Tweedie, F. J. (2001). Using Markov Chains for
Identification of Writer. Literary and Linguistic Computing 16 (3):
299–307. \\ https://doi.org/10.1093/llc/16.3.299.

\item [] Malladi, R.K. \& Prakash L.D. (2021). Time Series Analysis of
Cryptocurrency Returns and Volatilities. \emph{Journal of Economics and
Finance} 45 (1): 75–94.

\item[] Nakamoto, S. (2008). Bitcoin: A Peer-to-Peer Electronic Cash
System. \emph{Decentralized Business Review}, 212–60.\\
https://papers.ssrn.com/sol3/papers.cfm

\item [] Nascimento, F., Santos, S., Silva
J., Alves, X. \& Ferreira, T. (2022).
Extracting Rules via Markov Chains for Cryptocurrencies Returns
Forecasting. \emph{Computational Economics}, 1–20.\\
https://doi.org/10.1007/s10614-022-10237-7.

\item [] Urquhart, A. (2017), Price clustering in bitcoin,\emph{ Economics letters}, 159, 145–148.\\
https://doi.org/10.1016/j.econlet.2017.07.035.

\item [] Vilela-Mendes, R., Lima, R. \& Araujo, T. (2002). A
Process-Reconstruction Analysis of Market Fluctuations. \emph{International
journal of Theoretical Applied Finance} 5 (08): 797–821.\\
http://doi.org/10.1142/S0219024902001730.

\end{itemize}

%\end{thebibliography}

\section*{Declarations}

\begin{itemize}
\item Funding \\
This article is part of the Strategic Project UIDB/05069/2020. The authors acknowledge financial Support from FCT – Funda\c{c}\~{a}o para a Ci\^{e}ncia e Tecnologia (Portugal).
\item Conflict of interest/Competing interests
The authors have no conflicts of interest to declare that are relevant to the content of this article.
\item Ethics approval: Not applicable
\item Consent to participate: Not applicable
\item Consent for publication: Not applicable
\item Availability of data and materials
Data are available at \texttt{coinmarket.com}
\item Code availability
Code will be available at a GitHub public repository.
\item Authors' contributions

T. Ara\'{u}jo :  Methodology, Software, Supervision, Writing-Reviewing and Editing.\\
P. Barbosa: Data mining, Software, Writing - Original draft preparation.
\end{itemize}

\end{document}